\documentclass[runningheads]{llncs}

\usepackage{amsmath} 
\usepackage{amssymb} 

\usepackage{graphicx} 

\usepackage{hyperref} 

\usepackage{verbatim} 

\usepackage{subcaption}
\usepackage{mwe}

\usepackage{tikz}
\usetikzlibrary{arrows, arrows.meta, automata, positioning, shapes, decorations.pathmorphing, shadows }
\tikzset{every picture/.style={line width=0.75pt}} 

\usepackage{color}
\usepackage{xcolor}
\definecolor{mygray}{rgb}{0.4,0.4,0.4}
\definecolor{myblue}{HTML}{3A849E}
\definecolor{mygreen}{rgb}{0,0.6,0}
\definecolor{myorange}{HTML}{B25A00}
\definecolor{mygreen}{HTML}{467E7E}
\definecolor{mymauve}{HTML}{AC134D}

\definecolor{backred}   {rgb}{1  ,  0.8,  0.8}
\definecolor{backgreen} {rgb}{0.8,  1  ,  0.8}
\definecolor{backblue}  {rgb}{0.8,  0.8,  1}

\newcommand{\cucomment}[1]{}

\newcommand{\code}[1]{{\texttt{\textcolor{mygray}{\small #1}}}}

\usepackage{listings}

\lstdefinestyle{customc}{
  language=C,
  basicstyle=\linespread{0.9}\ttfamily\footnotesize,
  breakatwhitespace=false,
  breaklines=true,
  keywordstyle=\bfseries\color{mymauve!70},
  commentstyle=\itshape\color{mygreen},
  stringstyle=\color{orange},
  emph = {add0, add1, add2, add3, map_npt, map_page_wrong, map_page,},
  emphstyle=\bfseries\color{myblue!90},%
  deletekeywords={get},
  escapeinside={<@}{@>},
  keepspaces=true,
  otherkeywords={64, uint, ...},
  tabsize=2,
}

\lstdefinestyle{customcpp}{%
    language=C++,
    backgroundcolor=\color{black!5}, 
    basicstyle=\linespread{0.9}\ttfamily\footnotesize,
    breakatwhitespace=false,
    breaklines=true,
    keywordstyle=\bfseries\color{mymauve!70},
    commentstyle=\itshape\color{mygreen},
    stringstyle=\color{orange},
    emph = {add0, add1, add2, add3, map_npt, map_page_wrong, map_page,},
    emphstyle=\bfseries\color{myblue!90},%
    deletekeywords={get},
    escapeinside={<@}{@>},
    keepspaces=true,
    otherkeywords={64, uint, ...},
    tabsize=4,
}

\lstdefinestyle{customasm}{
    language=C,
    belowcaptionskip=1\baselineskip,
    abovecaptionskip=0pt,
    aboveskip=0pt,
    xleftmargin=\parindent,
    basicstyle=\footnotesize\ttfamily,
    commentstyle=\itshape\color{mygreen},
    keywordstyle=\bfseries\color{myblue!90},%
    emphstyle=\bfseries\color{myblue!90},%
    emph = {ld, st,dmb,cbz,ldax,cbnz,stx,cnbz,str,in},
    escapeinside={<@}{@>},
    breaklines=true,
    tabsize=2,
    boxpos=t,
    numberstyle=\footnotesize\color{gray}\tt,
}

\newcommand{\R}{\mathbb{R}}
\newcommand{\paren}[1]{\left(#1\right)}

\begin{document}

\title{FastAD:\@ 
    Expression Template-Based
    C++ Library for
    Fast and Memory-Efficient
    Automatic Differentiation}

\author{James Yang}
\institute{%
    Department of Statistics, Stanford University
    \email{jy2816@stanford.edu}
}
\authorrunning{J. Yang}
\titlerunning{FastAD:\@ C++ Template Library for AD}

\maketitle

\begin{abstract}
Automatic differentiation is a set of techniques to efficiently and accurately
compute the derivative of a function represented by a computer program.
Existing C++ libraries for automatic differentiation (e.g. Adept, Stan Math Library),
however, exhibit large memory consumptions and runtime performance issues.
This paper introduces FastAD, a new C++ template library for automatic differentiation,
that overcomes all of these challenges in existing libraries by using vectorization,
simpler memory management using a fully expression-template-based design,
and other compile-time optimizations to remove some run-time overhead.
Benchmarks show that FastAD performs 2-10 times faster than Adept
and 2-19 times faster than Stan
across various test cases including a few real-world examples.

\end{abstract}

\section{Introduction}

Gradient computation plays a critical role in modern computing problems 
surrounding optimization, statistics, and machine learning.
For example, one may wish to compute sensitivities of 
an Ordinary-Differential-Equation (ODE) integrator
for optimization or parameter estimation~\cite{carpenter:2015}.
In Bayesian statistics, advanced Markov-Chain-Monte-Carlo (MCMC) algorithms
such as the Hamiltonian Monte Carlo (HMC) and the No-U-Turn-Sampler (NUTS) rely
heavily on computing the gradient of a (log) joint probability density function
to update proposal samples in the leapfrog algorithm~\cite{hoffman:2011}\cite{neal:2012}.
Neural networks rely on computing 
the gradient of a loss function during back-propagation
to update the weights between each layer of the network~\cite{goodfellow:2016}.

Oftentimes, the target function to differentiate is extremely complicated,
and it is very tedious and error-prone for the programmer to manually define 
the analytical formula for the gradient~\cite{margossian:2018}.
It is rather desirable to have a generic framework where the programmer 
only needs to specify the target function to differentiate it.
Moreover, computing the derivative may be one of the more expensive parts of an algorithm
if it requires numerous such evaluations, as it is the case for the three examples discussed above.
Hence, it is imperative that gradient computation is as efficient as possible.
These desires motivated the development of such a framework: \textbf{automatic differentiation} (AD).

FastAD is a general-purpose AD library in C++ supporting both forward and reverse modes of automatic differentiation.
It is highly optimized to compute gradients, but also supports computing the full Jacobian matrix.
Similar to the Stan Math Library, FastAD is primarily intended for differentiating scalar functions, 
and it is well-known that reverse-mode is more efficient than forward-mode for these cases~\cite{carpenter:2015}.
For these reasons, this paper will focus only on reverse-mode and computing gradients 
rather than forward-mode or computing a full Jacobian matrix.

\section{Overview}

Section~\ref{sec:reverse} will first explain the reverse-mode automatic differentiation algorithm
to give context for how FastAD is implemented.
With this background, Section~\ref{sec:fastad} will discuss some implementation detail
and draw design comparisons with other libraries 
to see how FastAD overcomes the common challenges seen in these libraries.
In Section~\ref{sec:benchmark}, we show a series of benchmarks with
other libraries including two real-world examples 
of a regression model and a stochastic volatility model.

\section{Reverse-Mode Automatic Differentiation}\label{sec:reverse}

We briefly discuss the reverse-mode automatic differentiation algorithm
for context and to motivate the discussion for the next sections.
For a more in-depth treatment, we direct the readers 
to~\cite{carpenter:2015}\cite{margossian:2018}\cite{griewank:2008}.

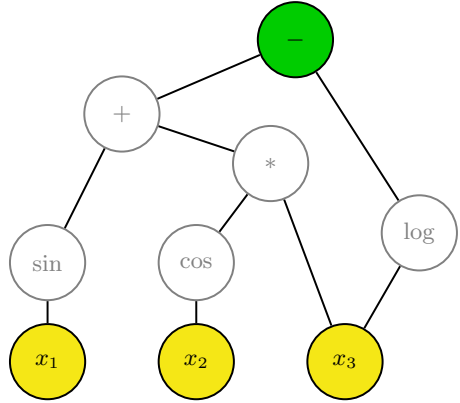
\begin{figure}[t]
\centering
\begin{tikzpicture}[x=0.75pt,y=0.75pt,yscale=-0.5,xscale=0.5]

    \draw (150,350) node [align=center, minimum size=1cm, draw, circle, fill=black!5!yellow] (x1)  {$x_1$};
    \draw (300,350) node [align=center, minimum size=1cm, draw, circle, fill=black!5!yellow] (x2)  {$x_2$};
    \draw (450,350) node [align=center, minimum size=1cm, draw, circle, fill=black!5!yellow] (x3)  {$x_3$};
    \draw (150,250) node [align=center, minimum size=1cm, draw, circle, color=gray] (sin) {$\sin$};
    \draw (300,250) node [align=center, minimum size=1cm, draw, circle, color=gray] (cos) {$\cos$};
    \draw (525,220) node [align=center, minimum size=1cm, draw, circle, color=gray] (log) {$\log$};
    \draw (225,100) node [align=center, minimum size=1cm, draw, circle, color=gray] (add) {$+$};
    \draw (375,150) node [align=center, minimum size=1cm, draw, circle, color=gray] (mul) {$*$};
    \draw (400,25)  node [align=center, minimum size=1cm, draw, circle, fill=black!20!green] (sub) {$-$};
    \draw (x1) -- (sin);
    \draw (x2) -- (cos);
    \draw (x3) -- (mul);
    \draw (x3) -- (log);
    \draw (sin) -- (add);
    \draw (cos) -- (mul);
    \draw (mul) -- (add);
    \draw (add) -- (sub);
    \draw (log) -- (sub);

\end{tikzpicture}

\caption{Expression graph for Eq.~\ref{eq:f-example}}\label{fig:expr-example}

\end{figure}

As an example, consider the function 
\begin{equation}
    f(x_1, x_2, x_3) = \sin(x_1) + \cos(x_2) \cdot x_3 - \log(x_3)
    \label{eq:f-example}
\end{equation}
This function can be represented as an expression graph
where each node of the graph represents a sub-expression.
Fig.~\ref{fig:expr-example} shows the corresponding expression graph
drawn to evaluate in the same order as defined by the operator precedence in the C++ standard.

\begin{figure}[t]
\centering
\begin{tikzpicture}[x=0.75pt,y=0.75pt,yscale=-0.5,xscale=0.5]

    \draw (150,450) node [align=center, minimum size=1cm, draw, circle, fill=gray] (x1)  {$x_1$};
    \draw (300,450) node [align=center, minimum size=1cm, draw, circle, fill=gray] (x2)  {$x_2$};
    \draw (450,450) node [align=center, minimum size=1cm, draw, circle, fill=gray] (x3)  {$x_3$};
    \draw (150,350) node [align=center, minimum size=1cm, draw, circle, fill=black!5!yellow] (w1)  {$w_1$};
    \draw (300,350) node [align=center, minimum size=1cm, draw, circle, fill=black!5!yellow] (w2)  {$w_2$};
    \draw (450,350) node [align=center, minimum size=1cm, draw=red, circle, fill=black!5!yellow] (w3)  {$w_3$};
    \draw (600,350) node [align=center, minimum size=1cm, draw=red, circle, fill=black!5!yellow] (w4)  {$w_4$};
    \draw (150,250) node [align=center, minimum size=1cm, draw, circle, color=gray] (sin) {$\sin$};
    \draw (300,250) node [align=center, minimum size=1cm, draw, circle, color=gray] (cos) {$\cos$};
    \draw (525,220) node [align=center, minimum size=1cm, draw, circle, color=gray] (log) {$\log$};
    \draw (225,100) node [align=center, minimum size=1cm, draw, circle, color=gray] (add) {$+$};
    \draw (375,150) node [align=center, minimum size=1cm, draw, circle, color=gray] (mul) {$*$};
    \draw (400,25)  node [align=center, minimum size=1cm, draw, circle, fill=black!20!green] (sub) {$-$};
    \draw [dashed] (w1) -- (x1);
    \draw [dashed] (w2) -- (x2);
    \draw [dashed] (w3) -- (x3);
    \draw [dashed] (w4) -- (x3);
    \draw (w1) -- (sin);
    \draw (w2) -- (cos);
    \draw (w3) -- (mul);
    \draw (w4) -- (log);
    \draw (sin) -- (add);
    \draw (cos) -- (mul);
    \draw (mul) -- (add);
    \draw (add) -- (sub);
    \draw (log) -- (sub);

\end{tikzpicture}

\caption{Converted expression tree for Eq.~\ref{eq:f-tree-example}.
         Nodes $x_1, x_2, x_3$ are now separated from the rest of the graph by a layer of $w$ variables.
         Note in particular that $x_3$ is now replaced with $w_3$ and $w_4$ (red boundary).
}\label{fig:expr-tree-example}

\end{figure}
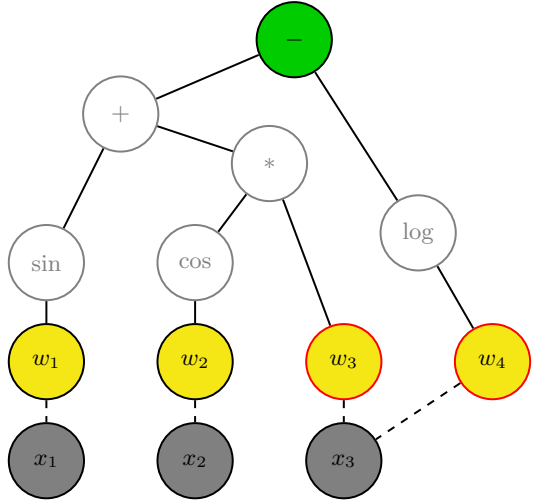

Note that, in general, a variable $x_i$ can be referenced by multiple nodes.
For example, $x_3$ is referenced by the \code{*} and \code{log} nodes.
It is actually more helpful to convert this expression graph into an expression tree
by replacing all such nodes with multiple parents as separate nodes
that have a reference back to the actual variable.
Mathematically,
\begin{align}
    f(x_1, x_2, x_3) &= \tilde{f}(g(x_1, x_2, x_3)) \label{eq:f-tree-example} \\
    \tilde{f}(w_1, w_2, w_3, w_4) &= \sin(w_1) + \cos(w_2) \cdot w_3 - \log(w_4) \nonumber \\
    g(x_1, x_2, x_3) &= (x_1, x_2, x_3, x_3) \nonumber
\end{align}
Fig.~\ref{fig:expr-tree-example} shows the corresponding converted expression tree.
This way all nodes except possibly the \code{$x_i$} nodes have exactly one parent
and have a unique path back up to the root, if we view the tree as a directed graph.
While this may seem more complicated than the original expression graph,
implementation becomes much cleaner with this new approach.
With the converted expression tree, we will see momentarily that 
we may actually start the algorithm from $w_i$ instead of $x_i$.
In fact, rather than treating $x_i$ as nodes of the expression graph,
it is more helpful to instead treat them as containers 
that hold the initial values and their \textbf{adjoints}, $\frac{\partial f}{\partial x_i}$.
For this reason, we denote the path between $x_i$ and $w_i$ with dotted lines,
and consider only up to nodes $w_i$ in the expression tree.

The reverse-mode algorithm consists of two passes of the expression graph:
\emph{forward}-evaluation (not to be confused with forward-mode AD), 
and \emph{backward}-evaluation.
During the \emph{forward}-evaluation, we compute the expression in the usual fashion,
i.e.\ start at the root, recursively forward-evaluate from left to right all of its children,
then finally take those results and evaluate the current node.
Evaluation of each node will compute the operation that it represents.
For example, after forward-evaluating the $w_1$ node, 
which is a trivial operation of retrieving the value from the container $x_1$,
the \code{sin} node will return~$\sin(w_1) = \sin(x_1)$.

The \emph{backward}-evaluation for a given node starts by receiving its adjoint from its parent.
We will also refer to this value as its \emph{seed} to further distinguish $x_i$ from the expression nodes.
Since the seed is exactly the partial derivative of $f$ with respect to the node,
the root of the expression tree will receive the value $ \frac{\partial f}{\partial f} = 1$.
Using the seed, the node computes the correct seed for all of its children and 
backward-evaluates from \emph{right-to-left}, the opposite direction of forward-evaluation.

The correct seed for each child is computed by a simple chain-rule.
Assuming the current node is represented by $w \in \R$ and one of its children is $v \in \R$,
the seed for $v$ is the following:
\begin{align}
    \frac{\partial f}{\partial v} &=
    \frac{df}{dw} \frac{\partial w}{\partial v} \label{eq:next-seed}
\end{align}
Each node $w$ has node-specific information to compute $\frac{\partial w}{\partial v}$.
For example, for the \code{log} node in Fig.~\ref{fig:expr-tree-example},
with $w$ as \code{log} and $v$ as $w_4$,
\begin{align*}
    \frac{\partial w}{\partial v} = \frac{d\log(w_4)}{dw_4} = \frac{1}{w_4}
\end{align*}
In general, if $v \in \R^{m \times n}$ and $w \in \R^{p \times q}$, then
\begin{align}
    \frac{\partial f}{\partial v_{ij}} &=
        \sum\limits_{k=1}^p \sum\limits_{l=1}^q 
        \frac{\partial f}{\partial w_{kl}} \frac{\partial w_{kl}}{\partial v_{ij}}
    \label{eq:next-adj}
\end{align}
In particular, the adjoint will always have the same shape and size as the value.

For nodes that have references back to the containers $x_i$, i.e.\ $w_1,\ldots,w_4$,
they must increment the adjoints in the containers with their seeds.
For example, nodes $w_3$ and $w_4$ will take their seeds and increment the adjoint for $x_3$.
This is easily seen by chain-rule again: let~$w_1, \ldots, w_k$ denote all of the variables that 
references $x$ and for simplicity assume they are all scalar, 
although the result can be easily generalized for multiple dimensions.
Then,
\begin{align*}
    \frac{\partial f}{\partial x} 
    &=  \sum\limits_{i=1}^k
        \frac{\partial f}{\partial w_{i}} \frac{\partial w_{i}}{\partial x}
    =   \sum\limits_{i=1}^k
        \frac{\partial f}{\partial w_{i}}
\end{align*}
where $\frac{\partial w_i}{\partial x} = 1$ because $w_i$ is simply the identity function with respect to $x$.
The fully accumulated adjoints for $x_1, x_2, x_3$ form the gradient of $f$.

In general, an expression node can be quite general and 
is only required to define how to compute its value and 
the adjoints of its children using Eq.~\ref{eq:next-adj}.

In the general case where $f: \R^n \to \R^m$,
we can apply this algorithm for the scalar functions $f_j$ for $j = 1,\ldots,m$ and
save each gradient as the $j$th row of a matrix.
The final matrix is then the Jacobian of $f$.

\section{FastAD Implementation}\label{sec:fastad}

In this section, we cover a few key ideas of our implementation\footnotemark.
\footnotetext{github page: https://github.com/JamesYang007/FastAD}
In Section~\ref{ssec:vectorization},
we first discuss the benefits of vectorization and the difficulties of integrating it into an AD system.
We then explain how FastAD fully utilizes vectorization
and demonstrate that other libraries do not fully take advantage of it. 
In Section~\ref{ssec:memory},
we discuss some memory management and performance issues
stemming from the use of the ``tape'' data structure.
We then explain how FastAD overcomes these challenges using expression templates
and a lazy allocation strategy.\@
Finally, Section~\ref{ssec:compile-time-opt} covers other compile-time optimizations 
that can further maximize performance.

\subsection{Vectorization}\label{ssec:vectorization}

Vectorization refers to the parallelization of operations on multiple data at the hardware level.
On a modern Intel 64-bit processor supporting AVX, 
four double-precision floating point numbers can be processed simultaneously,
roughly improving performance by a factor of four.
While the compiler optimization is able to vectorize a user's code sometimes, it is not guaranteed
because vectorization requirements are quite stringent. 
For example, vectorization is not guaranteed if memory access is not done in a contiguous fashion
and is impossible if there is any dependency between loop iterations.
This makes it quite challenging to design an AD system that 
can always predict compiler optimization to create vectorized code.
However, vectorization can make AD extremely fast, powerful, and practical even in complex problems.
In practice, we come across many examples where operations can be vectorized during gradient computation.
For example, matrix multiplication, any reduction from a multi-dimensional variable to a scalar such as
summation or product of all elements, and any unary and binary function that is applied element-wise such as
exponential, logarithm, power, sin, cos, tan, and the usual arithmetic operators.

Since most of the opportunities for vectorization occur in matrix operations,
the goal is to use a well-polished and efficient
matrix library such as \code{Eigen}, one of the most popular C++ matrix libraries.
However, this is not an easy task.
Adept2.0 notes in their documentation that integrating an expression-template based matrix library 
such as \code{Eigen} into an AD system can be quite challenging in design.
To circumvent these design issues, 
Adept2.0 integrates their own matrix library designed specifically for their AD system.
This, however, only introduces another problem of writing an efficient matrix library, another daunting task.
In fact, the author of Adept notes that matrix multiplication,
one of the most widely used matrix operations, 
is currently very slow~\cite{hogan:2014}.
Stan mentions that integrating matrix libraries can lead to unexpected problems 
that would require extra memory allocations to resolve~\cite{carpenter:2015}.
Other libraries such as ADOL-C, CppAD, and Sacado do not integrate any matrix libraries directly,
but they do allow the use of \code{Eigen} matrix classes simply as containers.

The one library among those mentioned previously that ultimately incorporates \code{Eigen} is Stan.
Stan provides their own ``plug-ins'', which extend \code{Eigen} classes for their AD variables.
For example, in Stan, one would use \\
\code{Eigen::Matrix<stan::math::var, \ldots>} as a matrix of AD (univariate) variables,
where the generic matrix object is extended to work differently for \code{stan::math::var}.
However, this design still does not take full advantage of the vectorization that \code{Eigen} provides.
One reason is that \code{Eigen} is only highly optimized for matrices of primitive types such as double-precision floating points and integers.
For any other class types, \code{Eigen} defaults to a less optimized version with no guarantees of vectorization.
Therefore, unless one extends all \code{Eigen} methods with optimizations for their specific class type,
they will not receive much benefit of vectorization.
Another reason is that these matrices now have a heterogeneous structure where
each element is an AD variable which represents a pair of value and adjoint.
As a result, it is not possible to read only the values (and similarly, only the adjoints) of the matrix in a contiguous fashion.
The compiler then cannot guarantee any automatic vectorization.

\begin{figure*}[t]
    \centering
    \includegraphics[width=0.8\textwidth]{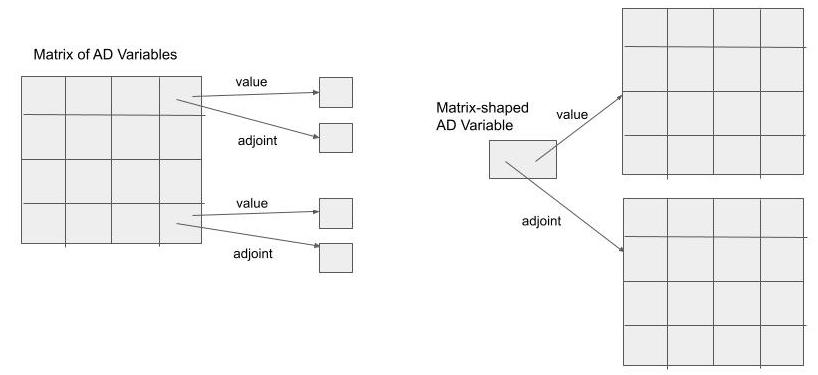}
    \caption{%
        The left shows the memory layout for a matrix of univariate AD variables.
        This refers to the design pattern ``a matrix of pair of pointers to double''
        since each element of the matrix contains two pointers pointing to its value and adjoint.
        The right shows the memory layout for a single matrix-shaped AD variable,
        referring to the reverse pattern ``a pair of pointers to matrix of doubles''.
        This AD variable has two pointers, each pointing to a matrix of doubles for the value and adjoint.
    }\label{fig:matrix-memory-layout}
\end{figure*}

FastAD fully utilizes the benefits of \code{Eigen} through one simple design difference, which we refer to as \emph{shape traits}.
Stan as well as the other libraries mentioned previously except Adept2.0
follow the design pattern of ``a matrix of pair of pointers to double'' when defining a matrix of AD variables.
Note that a univariate AD variable internally contains two pointers to doubles,
one pointing to the value and one to the adjoint.
In contrast, FastAD follows the reverse pattern: ``a pair of pointers to matrix of doubles''.
That is, rather than defining a univariate AD variable, which gets reused as an element of a matrix,
we define a separate matrix AD variable containing a pair of pointers, each pointing to a matrix of double.
Figure~\ref{fig:matrix-memory-layout} shows the memory layout for each of the design patterns.
This subtle difference provides a few important benefits.
Since the value and adjoint are now represented as matrices of primitive types,
matrix operations will be vectorized.
The second benefit is the significant reduction of memory consumption.
For other libraries, a matrix of AD variables contains two pointers \emph{for each element}.
However, with our design, a single matrix AD variable would only store two pointers
regardless of the size of the matrix.
Hence, if a matrix is $m\times n$, 
the traditional approach has a $O(mn)$ memory consumption for the pointers,
and FastAD approach has a $O(1)$ consumption.

Using templates in C++,
it is easy to unify the API for the different AD variable shapes
by providing an additional template parameter:
\begin{lstlisting}[style=customcpp]
ad::Var<double, ad::scl> x; // scalar shape
ad::Var<double, ad::vec> v; // vector shape
ad::Var<double, ad::mat> m; // matrix shape
\end{lstlisting}

Shape traits also apply for any arbitrary node because
we can deduce the output shape given the input shapes.
The following is a declaration of a generic node representing an
element-wise unary operation, which demonstrates this idea:
\begin{lstlisting}[style=customcpp]
template <class Unary, class ExprType>
struct UnaryNode:
    ValueAdjView<
      typename util::expr_traits<ExprType>::value_t,
      <@\textcolor{red}{typename util::shape\_traits<ExprType>::shape\_t}@>>
{ /*...*/ };
\end{lstlisting}
The portion highlighted in red related to \code{shape\_t}
takes the input expression type \code{ExprType} and deduces its shape type.
Since an element-wise unary operation always has the same shape as its input,
the unary node takes on the same shape.

To verify that FastAD vectorizes more than Stan, 
we performed reverse-AD for a simple summation function $f(x) = \sum\limits_{i=1}^n x_i$
and generated the disassembly for Stan and FastAD~\footnotemark.\@
\footnotetext{github page: https://github.com/JamesYang007/ADBenchmark}
We extracted the part that performs the summation
and compared the instructions to see whether vectorization was taking place.\@
The following is the disassembly for Stan:
\begin{lstlisting}[style=customasm]
L3178:
    movq    (%rax), %rdx
    addq    $8, %rax
    <@\textcolor{red}{vaddsd}@>   8(%rdx), %xmm0, %xmm0 
    cmpq    %rcx, %rax 
    jne L3178
\end{lstlisting}
The instruction used to add is \code{vaddsd},
which is an AVX instruction to add \emph{scalar} double-precision values.
This is not a vectorized instruction, and hence the addition is not done in parallel.
This portion of the disassembly is related to a specialization of an \code{Eigen} class 
responsible for summation with \emph{default traversal},
which is no different from a naive for-loop.

Compare the above disassembly with the one generated for FastAD:
\begin{lstlisting}[style=customasm]
 L3020:
     addq    $8, %rdx
     <@\textcolor{red}{vaddpd}@>  (%rax), %ymm1, %ymm1   
     <@\textcolor{red}{vaddpd}@>  32(%rax), %ymm0, %ymm0 
     addq    $64, %rax
     cmpq    %rdx, %rcx 
     jg  L3020 
\end{lstlisting}
This portion of the assembly is indeed related to the \emph{linear vectorized traversal}
specialization of the same \code{Eigen} class responsible for the summation.
The instruction used to add is \code{vaddpd},
which is an AVX instruction to add \emph{packed} double-precision values.
This is a vectorized instruction and the operation is truly done in parallel.

Sometimes, Stan is able to produce vectorized code such as in matrix multiplication.
This is consistent with our benchmark results 
since Stan came closest to FastAD for this operation (see Section~\ref{ssec:matrix_mult}).
It is also consistent with how it is implemented,
since they allocate extra memory for \code{double} values for each matrix 
and the multiplication is carried out with these matrices of primitive types.
However, this vectorization does come at a cost of at least 4 times extra memory allocation than what FastAD allocates.
Moreover, the backward-evaluation requires heap-allocating a matrix on-the-fly every time.
FastAD incurs no such cost, only allocates what is needed, and never heap-allocates during AD evaluation.

\subsection{Memory Management}\label{ssec:memory}

Most AD systems manage a data structure in memory often referred to as the ``tape''
to store the sequence of operations via function pointers as well as the node values and adjoints.
This tape is modified dynamically and requires sophisticated memory management 
to efficiently reuse memory whenever possible.
Stan even writes their own custom memory allocator to alleviate memory fragmentation,
promote data locality, and amortize the cost of memory allocations~\cite{carpenter:2015}.
However, the memory is not fully contiguous and may still over-allocate.
For some libraries, on top of memory management of these operations,
a run-time check must be performed at every evaluation to determine the correct operation~\cite{bell:2020}.
Others like Stan rely on dynamic polymorphism to look up the vtable to call the correct operation~\cite{carpenter:2015}.

FastAD is unique in that it uses expression templates to represent the sequence of operations
as a single stack-allocated object.
By overloading the comma operator, we can chain expressions together into a single object.
The following is an example of chaining multiple expressions:
\begin{lstlisting}[style=customcpp]
auto expr = (
    x = y * z,                      // expr 1
    w = x * x + 3 * ad::sin(x + y), // expr 2
    w + z * x                       // expr 3
);
\end{lstlisting}
Each of the three sub-expressions separated by the commas returns an expression object
containing the necessary information to perform reverse-AD on their respective structure.
Those expression objects are then chained by the comma operators to build a final expression object
that contains the information to perform reverse-AD on all 3 sub-expressions in the order presented.
This final object is saved into the variable \code{expr}.

Expression template makes it possible to build these expression objects containing the reverse-AD logic.
Expression template is a template metaprogramming technique that builds complex
structures representing a computation at compile-time.
For a full treatment of expression templates, we direct the readers to~\cite{vandevoorde:2002}.
As an example, in the following case,
\begin{lstlisting}[style=customcpp]
Var<double, scl> x, y;
auto expr = x + y;
\end{lstlisting}
\code{x+y} returns a new object of type
\code{BinaryNode<Add, Var<double, scl>, Var<double, scl>>},
which represents the addition of two AD variables.
In particular, this object has member functions that define the logic for
the forward and backward evaluations of the reverse-mode AD.\@
This design brings many performance benefits.
Since the compiler now knows the entire sequence of operations for the expression,
it allows for the reverse-AD logic to be inlined with no virtual function calls,
and it removes the need to dynamically manage a separate vector of function pointers.
Additionally, the expression object is stack-allocated,
which is cheaper than being heap-allocated,
and its byte size is proportional to the number of nodes,
which is relatively small in practice.

We can optimize the memory management even further with another observation:
an expression can determine the \emph{exact} number of values and adjoints needed to represent all of the nodes.
If every node can determine this number for itself,
then any expression tree built using these nodes can determine it as well by induction.
It is the case that all nodes defined in FastAD can, in fact, determine this number.
This leads to the idea of \emph{lazy allocation}.
Lazy allocation refers to allocating memory only when the memory is required by the user.
In other words, an expression object does not necessarily need to 
allocate memory for the values and adjoints at construction time,
since it is only needed later at differentiation time.
Once an expression is fully built and the user is ready to differentiate it,
the user can lazily determine this number of values and adjoints by interfacing with the expression object,
allocate that amount in a contiguous manner,
and ``bind'' the expression object to use that region of memory.
This solves the issue with the traditional tape where the values and adjoints are
not stored in a fully contiguous manner.
Conveniently, the allocated values and adjoints also do not necessarily 
need to be stored and managed by a global tape.
Furthermore, the expression objects can be given additional logic 
to bind to this region systematically so that the forward and backward evaluations
will read this memory region almost linearly, which increases cache hits.

While Stan notes that they are more memory-efficient than other popular C++ libraries~\cite{carpenter:2015},
we noticed a non-negligible difference in memory consumption between Stan and FastAD.
We took the stochastic volatility example in Section~\ref{ssec:stochastic_volatility},
and compared the memory allocation in number of bytes.
For Stan, we took the member variables
from the tape and computed the number of used bytes.
We did not take into account other miscellaneous members for simplicity,
and this estimate serves as a very rough lower bound on the total amount of memory allocated.
For FastAD, we computed the memory allocated for the values, adjoints, and the expression object.
Our result shows that Stan uses 4718696 bytes and FastAD uses 1836216 bytes.
This rough estimate shows that Stan uses at least 2.5 times more memory than FastAD.

With all these optimizations, FastAD removes the need for a global tape by 
using expression templates to manage the sequence of operations in an expression at compile-time,
contiguously allocating the exact number of values and adjoints,
and localizing the memory allocations for each expression.

\subsection{Other Compile-Time Optimizations}\label{ssec:compile-time-opt}

Using C++17 features, we can make further compile-time optimizations
that could potentially save tremendous amount of time during run-time.


One example is choosing the correct specialization of an operation
depending on the shapes of the input.
As seen in Section~\ref{ssec:vectorization}, all nodes are given a shape trait.
Depending on the input shapes, one may need to invoke different routines for the same node.
For example, the normal log-pdf node behaves quite differently
depending on whether the variance parameter is a scalar $\sigma^2$
or a (covariance) matrix $\Sigma$.
Namely, if the variance has a matrix shape,
we must perform a matrix inverse to compute the log-pdf,
which requires a different code from the scalar case.
Using a C++ design pattern called Substitution-Failure-Is-Not-An-Error (SFINAE),
we can choose the correct routine at compile-time.
The benefit is that there is no time spent during run-time in choosing the routine anymore,
whereas in libraries like CppAD, they choose the routines at run-time 
\emph{for every evaluation} of the node~\cite{bell:2020}.


Another example is detecting constants in an expression. 
We can optimize a node by saving temporary results when certain inputs are constants,
which we can check at compile-time using the C++ language feature \code{if constexpr}.
These results can then be reused in subsequent AD evaluations,
sometimes changing orders of magnitude of the performance.
As an example, consider an expression containing a normal log-pdf node for which we differentiate many times.
Suppose also that the input variables to the node are $x$, an $n$-dimensional vector of constants,
and $\mu, \sigma$, which are scalar AD variables.
In general, for every evaluation of the node, the time to forward-evaluate the log-pdf is $O(n)$,
since we must compute 
\begin{align*}
    \log(p(x|\mu, \sigma)) &= -\frac{\sum\limits_{i=1}^n (x_i-\mu)^2}{2\sigma^2} - n\log(\sigma) + C
\end{align*}
However, the normal log-pdf node has an opportunity to make a powerful optimization in this particular case.
Since the normal family defines a two-dimensional exponential family,
it admits a sufficient statistic $T(x) = \paren{\bar{x}, S^2_n(x)}$ where
\begin{align*}
    \bar{x} := \frac{1}{n} \sum\limits_{i=1}^n x_i, \,
    S_n^2(x) := \frac{1}{n} \sum\limits_{i=1}^n (x_i - \bar{x})^2
\end{align*}
Since $x$ is a constant, the sufficient statistic can then be computed \emph{once} and saved for future use.
The normal log-pdf forward-evaluation now only takes $O(1)$ time given the sufficient statistic,
as seen in Eq.~\ref{eq:normal_log_pdf_x_const} below,
\begin{align}
    \log(p(x|\mu, \sigma)) 
    &= -\frac{1}{2 \sigma^2} 
        \sum\limits_{i=1}^n (x_i - \mu)^2 
        - n\log(\sigma) + C \nonumber \\
    &= -\frac{n}{2 \sigma^2} 
        \paren{S_n^2 + (\bar{x} - \mu)^2} 
        - n\log(\sigma) + C 
        \label{eq:normal_log_pdf_x_const}
\end{align}

\section{Benchmarks}\label{sec:benchmark}

In this section, we compare performances of 
various libraries against FastAD for a range of examples\footnotemark.
\footnotetext{github page: https://github.com/JamesYang007/ADBenchmark}
The following is an alphabetical list of the libraries used for benchmarking:\@
\begin{itemize}
    \item \href{http://www.met.reading.ac.uk/clouds/adept/}{Adept 2.0.8}~\cite{hogan:2014}
    \item \href{https://github.com/coin-or/ADOL-C}{ADOL-C 2.7.2}~\cite{griewank:1996}
    \item \href{https://coin-or.github.io/CppAD/doc/cppad.htm}{CppAD 20200000}~\cite{bell:2020}
    \item \href{https://github.com/JamesYang007/FastAD}{FastAD 3.1.0}
    \item \href{https://github.com/trilinos/Trilinos/tree/master/packages/sacado}{Sacado 13.0.0}~\cite{phipps:2009}
    \item \href{https://github.com/stan-dev/math}{Stan Math Library 3.3.0}~\cite{carpenter:2015}
\end{itemize}
All the libraries are at their most recent release at the time of benchmarking.
These libraries have also been used by others~\cite{carpenter:2015}\cite{margossian:2018}\cite{hogan:2014}.

All benchmarks were run on a Google Cloud Virtual Machine with the following configuration:
\begin{itemize}
    \item \textbf{OS}: Ubuntu 18.04.1 
    \item \textbf{Architecture}: x86 64-bit
    \item \textbf{Processor}: Intel Xeon Processor 
    \item \textbf{Cores}: 8
    \item \textbf{Compiler}: GCC10
    \item \textbf{C++ Standard}: 17
    \item \textbf{Compiler Optimization Flags}: \code{-O3 -march=native} 
\end{itemize}

All benchmarks benchmark the case where a user wishes to differentiate
a scalar function $f$ for different values of $x$.
This is a very common use-case.
For example, in the Newton-Raphson method,
we have to compute $f'(x_n)$ at every iteration with the updated $x_n$ value.
In HMC and NUTS, the leapfrog algorithm frequently
updates a quantity called the ``momentum vector'' $\rho$ 
with $\nabla_\theta \log(p(\theta, x))$ ($x$ here is treated as a constant),
where $\theta$ is a ``position vector'' that also gets frequently updated.

Our benchmark drivers are very similar to the ones used by Stan~\cite{carpenter:2015}.
As noted in~\cite{margossian:2018}, there is no standard set of benchmark examples for AD,
but since Stan is most similar to FastAD in both design and intended usage,
we wanted to keep the benchmark as similar as possible.

We measure the average time to differentiate a function
with an initial input and 
save the function evaluation result as well as the gradient.
After timing each library, the gradients are compared with manually-written gradient computation to check accuracy.
All libraries had some numerical issues for some of the examples,
but the maximum proportion of error to the actual gradient values was on the order of $ 10^{-15}$, which is negligible.
Hence, in terms of accuracy, all libraries were equally acceptable.
Finally, all benchmarks were written in the most optimal way for every library based on their documentation.

Every benchmark also times the ``baseline'', 
which is a manually-written forward evaluation (computing function value).
This will serve as a way to measure the extra overhead of computing the gradient relative to computing the function value.
This approach was also used in~\cite{carpenter:2015},
however in our benchmark, the baseline is also optimized to be vectorized.
Hence, our results for all libraries with respect to the baseline may differ from those in the reference.

Sections~\ref{ssec:sum_prod}-\ref{ssec:normal_log_pdf} 
cover some micro-benchmarks where we benchmark commonly used functions: 
summation and product of elements in a vector, 
log-sum-exponential, 
matrix multiplication, 
and normal log-pdf.
Sections~\ref{ssec:regression}-\ref{ssec:stochastic_volatility} 
cover some macro-benchmarks where we benchmark practical examples: 
a regression model and a stochastic volatility model.

\subsection{Sum and Product}\label{ssec:sum_prod}

\begin{figure*}[t]
    \centering
    \begin{subfigure}[b]{0.475\textwidth}
        \centering
        \includegraphics[width=\textwidth]{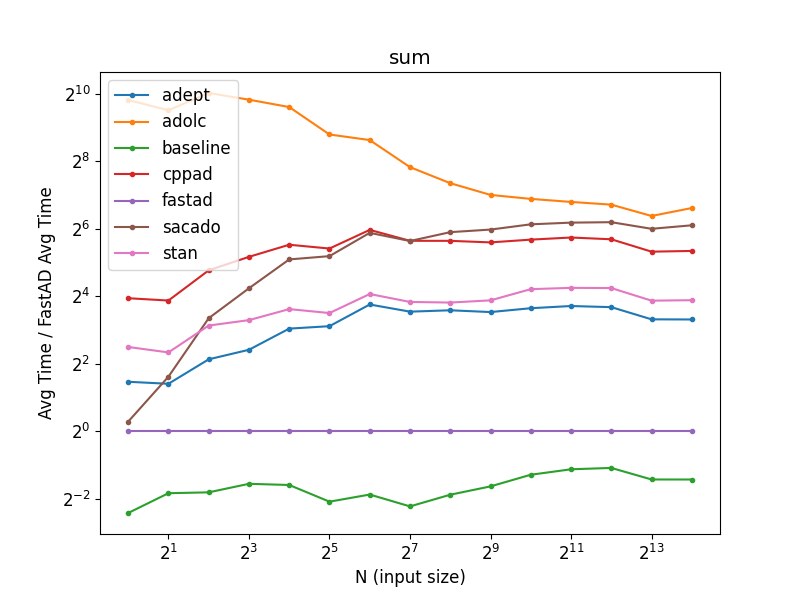}
        \caption{Sum}\label{fig:sum}
    \end{subfigure}
    \hfill
    \begin{subfigure}[b]{0.475\textwidth}
        \centering
        \includegraphics[width=\textwidth]{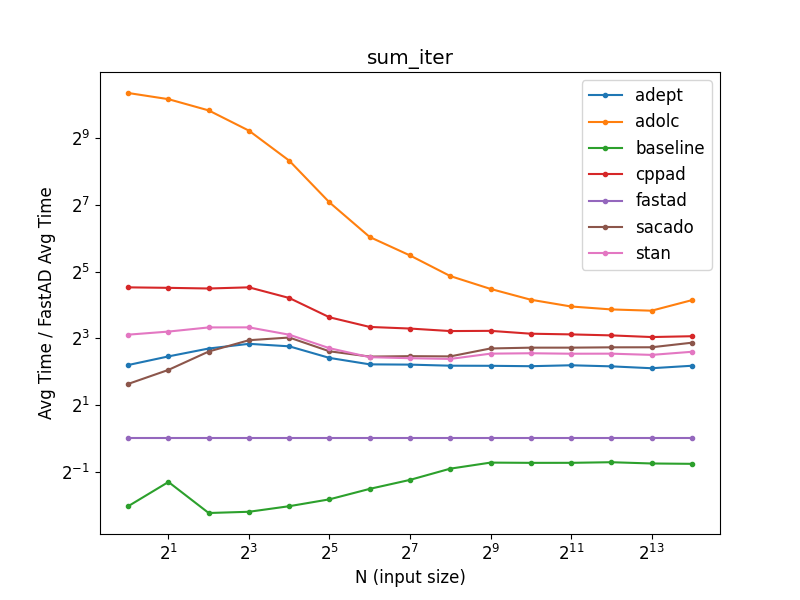}
        \caption{Sum (naive, for-loop-based)}\label{fig:sum_iter}
    \end{subfigure}
    \vskip\baselineskip%
    \begin{subfigure}[b]{0.475\textwidth}
        \centering
        \includegraphics[width=\textwidth]{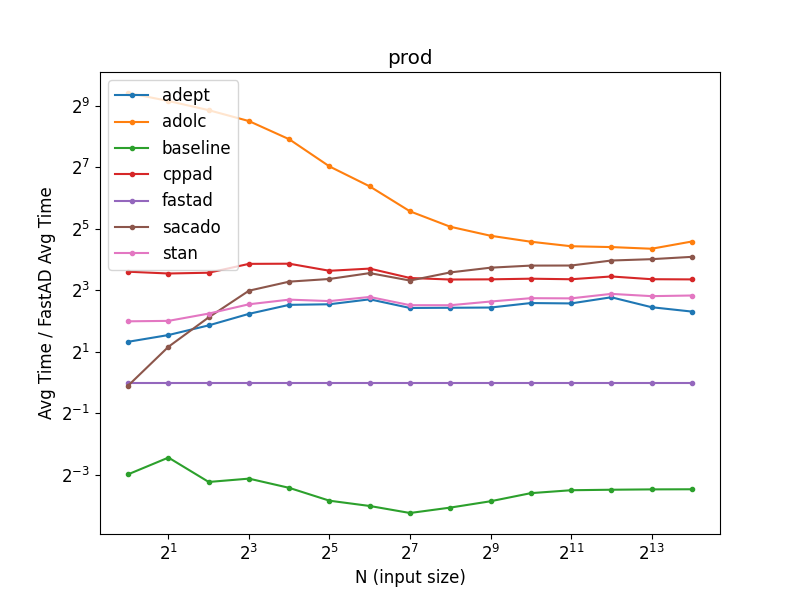}
        \caption{Product}\label{fig:prod}
    \end{subfigure}
    \hfill
    \begin{subfigure}[b]{0.475\textwidth}
        \centering
        \includegraphics[width=\textwidth]{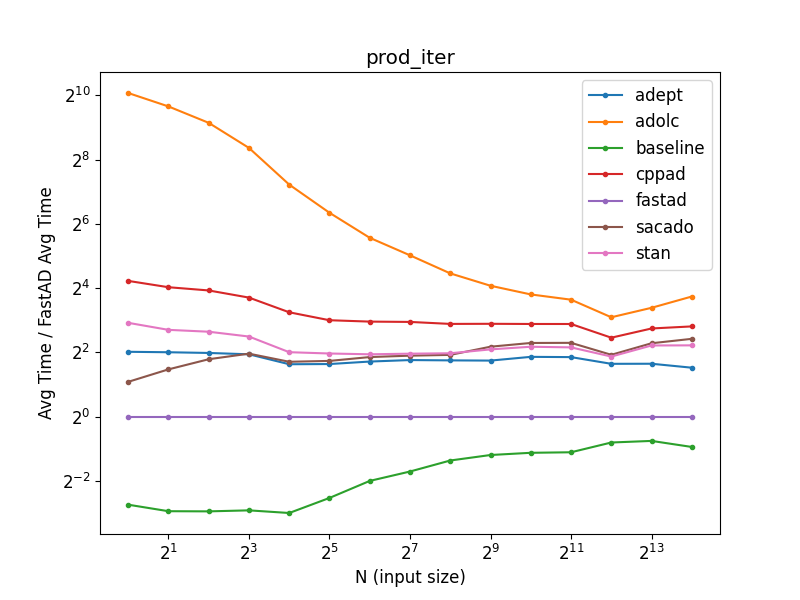}
        \caption{Product (naive, for-loop-based)}\label{fig:prod_iter}
    \end{subfigure}
    \caption{%
        Sum and product benchmarks of other libraries against FastAD 
        plotted relative to FastAD average time.
        Fig.~\ref{fig:sum},\ref{fig:prod} use built-in functions whenever available.
        Fig.~\ref{fig:sum_iter},\ref{fig:prod_iter} use the naive iterative-based code for all libraries.
    }\label{fig:sum_prod}
\end{figure*}

The summation function is defined as $f_s:\R^n \to \R$ where~$f_s(x) = \sum\limits_{i=1}^n x_i$,
and the product function is defined as $f_p:\R^n \to \R$ where~$f_p(x) = \prod\limits_{i=1}^n x_i$.
We also tested the case where we forced all libraries to use a manually-written for-loop-based summation and product.
Fig.~\ref{fig:sum_prod} shows the benchmark results.

In all four cases, it is clear that FastAD outperforms all libraries for all values of $N$
by a significant factor.
For \code{sum}, 
the next three fastest libraries, asymptotically, are Adept, Stan, and CppAD, respectively.
The trend stabilizes starting from $N=2^6=64$ where Adept is about $ 10$ times slower than
FastAD and Stan about $ 15$ times.
The naive, for-loop-based benchmark shown in Fig.~\ref{fig:sum_iter} exhibits a similar pattern,
although the performance difference with FastAD is less extreme:
Adept is about $ 4.5$ times slower
and Stan about $ 6$ times.
Nonetheless, this is still a very notable difference.

For the \code{prod} and \code{prod\_iter} cases, 
Fig.~\ref{fig:prod},\ref{fig:prod_iter} again show
a similar trend as \code{sum} and \code{sum\_iter}.
Overall, the comparison with FastAD is less extreme.
For \code{prod}, Adept is about $ 5$ times slower than FastAD,
and Stan about $ 7$ times.
For \code{prod\_iter}, Adept is about $3$ times slower
and Stan about $4.7$ times.

\subsection{Log-Sum-Exp}

\begin{figure*}[t]
    \centering
    \includegraphics[width=0.8\textwidth]{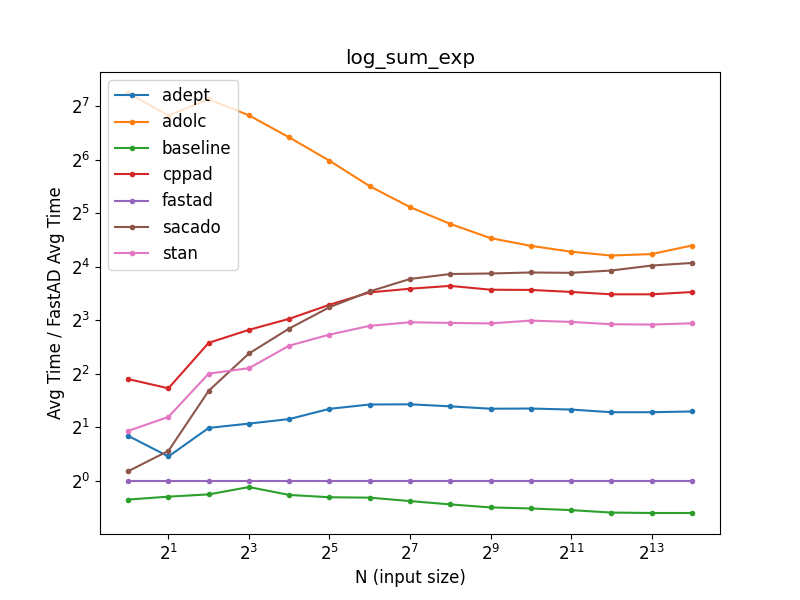}
    \caption{%
        Log-sum-exp benchmark of other libraries against FastAD 
        plotted relative to FastAD average time.
    }\label{fig:log_sum_exp}
\end{figure*}

The log-sum-exp function is defined as $f:\R^n \to \R$ 
where~$f(x) = \log(\sum\limits_{i=1}^n e^{x_i})$,
Fig.~\ref{fig:log_sum_exp} shows the benchmark results.

FastAD outperforms all libraries for all values of $N$.
The next three fastest libraries are Adept, Stan, and CppAD, respectively.
The trend stabilizes starting from $N=2^6=64$ where 
Adept is about $ 2.5$ times slower than FastAD, 
and Stan about $ 7.7$ times.
Note that FastAD is only marginally slower than the baseline,
especially for small to moderate values of $N$,
despite calls to expensive functions like \code{log} and \code{exp}.
In fact, at the largest value of $N = 2^{14}$, 
FastAD is $ 1.5$ times slower than the baseline, 
meaning there is about $ 50\%$
overhead from one forward-evaluation to also compute the gradient.
This is because FastAD is optimized such that in this case
the forward-evaluated result for \code{exp} expression is reused
during the backward-evaluation.

\subsection{Matrix Multiplication}\label{ssec:matrix_mult}

\begin{figure*}[t]
    \centering
    \includegraphics[width=0.8\textwidth]{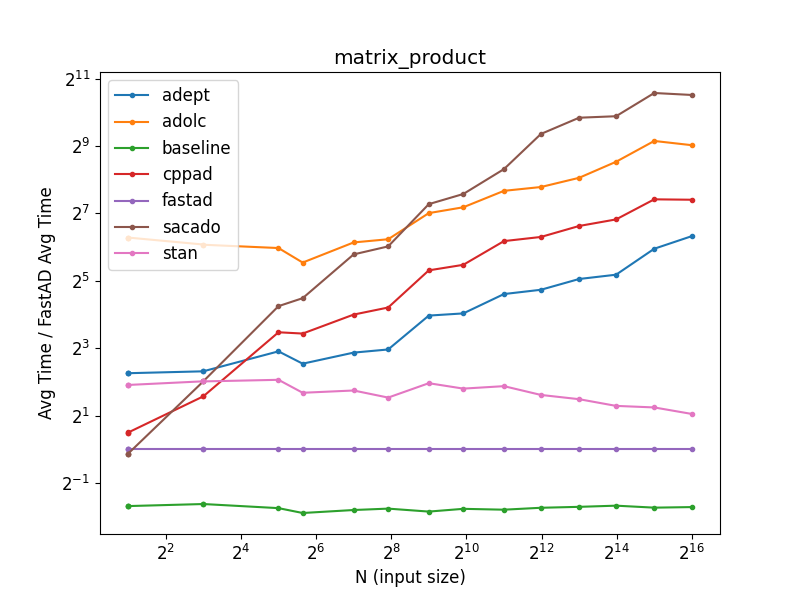}
    \caption{%
        Matrix multiplication benchmark of other libraries against FastAD 
        plotted relative to FastAD average time.
    }\label{fig:matrix_mult}
\end{figure*}

For this benchmark, all matrices are square matrices of the same size.
In order to have a scalar target function,
we add another step of adding all of the entries of the matrix multiplication, i.e.
\[
    f(A, B) = \sum\limits_{i=1}^{K} \sum\limits_{j=1}^{K} {(A \cdot B)}_{ij}
\]
Fig.~\ref{fig:matrix_mult} shows the benchmark results.

FastAD is still the fastest library for all values of $N$, 
but Stan performs much closer to FastAD than in the previous examples.
All other libraries consistently take longer than both FastAD and Stan as $N$ increases.
Towards the end at around $N=2^{14}$, 
Stan is about $ 2$ times slower.
For moderate sized $N \in [2^{8}, 2^{10}]$, Stan is about $ 3$ times slower.

This example really highlights the benefits of vectorization.
As noted in Section~\ref{ssec:vectorization},
this was the one benchmark example where Stan was able to produce vectorized code,
which is consistent with Figure~\ref{fig:matrix_mult} 
that Stan is the only library that has the same 
order of magnitude as FastAD.
Other libraries did not produce vectorized code.

The comparison with the baseline shows that FastAD takes $ 3.27$ times longer.
Note that forward-evaluation requires one matrix multiplication between two $K\times K$ matrices,
and backward-evaluation additionally requires two matrix multiplications of the same order,
one for each adjoint:
\begin{align*}
    \frac{\partial f}{\partial A} 
    &= \frac{\partial f}{\partial (A\cdot B)} \cdot B^T, \,
    \frac{\partial f}{\partial B} 
    = A^T \cdot \frac{\partial f}{\partial (A\cdot B)}
\end{align*}
Hence, in total, one AD evaluation requires three matrix multiplications between two $K\times K$ matrices.
If we approximate a manually-written gradient computation to take 
three times as long as the baseline (one multiplication), 
FastAD time relative to this approximated time
is $\frac{3.27}{3} = 1.09$.
This shows then that FastAD only has about $ 9\%$ overhead 
from a manually-written code, which is extremely optimal.

\subsection{Normal Log-PDF}\label{ssec:normal_log_pdf}

\begin{figure*}[t]
    \centering
    \includegraphics[width=0.7\textwidth]{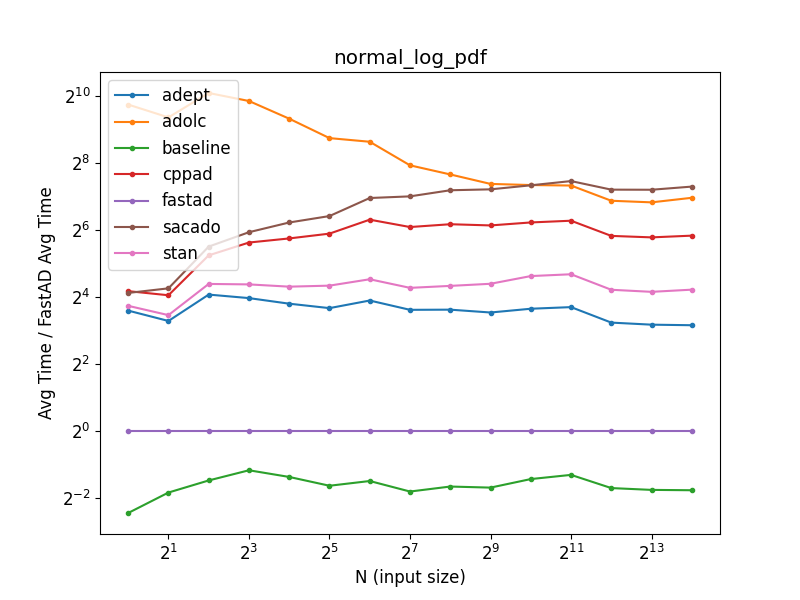}
    \caption{%
        Normal log-pdf benchmark of other libraries against FastAD 
        plotted relative to FastAD average time.
    }\label{fig:normal_log_pdf}
\end{figure*}

The normal log-pdf function is defined up to a constant as:
\[
    f(x) = -\frac{1}{2\sigma^2} \sum\limits_{i=1}^N \paren{x_i - \mu}^2 
           -N\log(\sigma)
\]
For this benchmark, $\mu = -0.56,\,\sigma = 1.37$ and are kept as constants.
Fig.~\ref{fig:normal_log_pdf} shows the benchmark results.

FastAD is the fastest library for all values of $N$.
The trend stabilizes at around $N=2^{7}=128$.
Towards the end, Adept is about $ 9$ times slower,
and Stan about $ 19$ times slower.
Comparing all libraries,
the overall difference we see in this example is the largest we have seen so far,
and this is partly due to how we compute $\log(\sigma)$.
Section~\ref{ssec:compile-time-opt} showed that we can check at compile-time
whether a certain variable is a constant, in which case,
we can perform a more optimized routine.
In this case, since $\sigma$ is a constant, it computes the normalizing constant
$\log(\sigma)$ once and gets reused over multiple AD evaluations 
with no additional cost during runtime,
which saves a lot of time since logarithm is a relatively expensive function.

\subsection{Bayesian Linear Regression}\label{ssec:regression}

\begin{figure*}[t]
    \centering
    \includegraphics[width=0.75\textwidth]{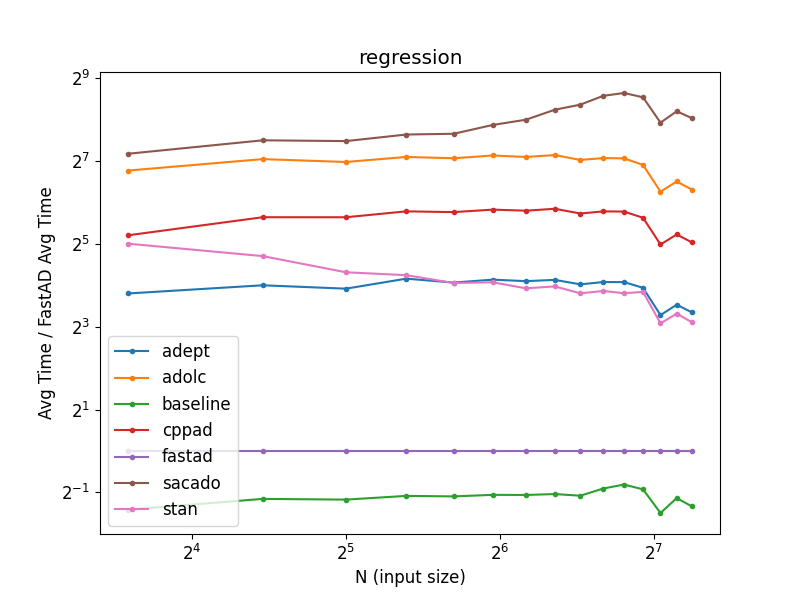}
    \caption{%
        Bayesian linear regression benchmark of Stan against FastAD 
        plotted relative to FastAD average time.
    }\label{fig:regression}
\end{figure*}

This section marks the first macro-benchmark example.
We consider the following Bayesian linear regression model:
\begin{align*}
    y &\sim N\paren{X\cdot w + b, \sigma^2} \\
    w &\sim N\paren{0,1} \\
    b &\sim N\paren{0,1} \\
    \sigma &\sim Unif\paren{0.1, 10.}
\end{align*}
The target function is the log of the joint probability density function (up to a constant).
Fig.~\ref{fig:regression} shows the benchmark results.

FastAD outperforms Stan by $ 8.6$ times for the largest $N$ and Adept by $ 10$ times.
The trend stabilizes starting from around $N=70$.
It is interesting to see that around $N=2^7$, 
FastAD is only $ 2.2$ times slower than the baseline,
despite the model consisting of a large matrix multiplication and many normal log-pdfs.
One of the reasons is that the compiler was able to optimize-out the backward-evaluation 
of the matrix constant $X$, since constants implement a no-op for backward-evaluation.

If we assume that the most expensive operation is the matrix multiplication,
AD evaluation approximately takes two matrix multiplications between a matrix and a vector.
We can then approximate a lower bound for the manually-written gradient computation time to be two times that of the baseline.
The relative time of FastAD to this approximated time is
$1.1$, implying about $ 10\%$ overhead from a manually-written code.

\subsection{Stochastic Volatility}\label{ssec:stochastic_volatility}

\begin{figure*}[t]
    \centering
    \includegraphics[width=0.8\textwidth]{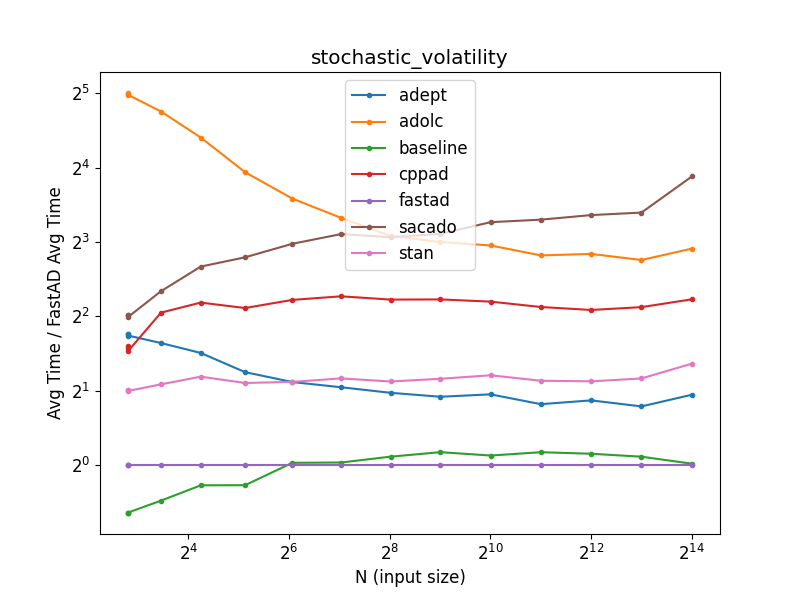}
    \caption{%
        Stochastic volatility benchmark of Stan against FastAD 
        plotted relative to FastAD average time.
    }\label{fig:stochastic_volatility}
\end{figure*}

This section marks the second and last macro-benchmark example.
We consider the following stochastic volatility model 
taken from the Stan user guide~\cite{stan-rm:2018}:
\begin{align*}
    y &\sim N(0, e^{h}) \\
    h_{std} &\sim N(0, 1) \\
    \sigma &\sim Cauchy(0,5) \\
    \mu &\sim Cauchy(0,10) \\
    \phi &\sim Unif(-1, 1) \\
    h &= h_{std} \cdot \sigma \\
    h[0] &= \frac{h[0]}{\sqrt{1 - \phi^2}} \\
    h &= h + \mu \\
    h[i] &= \phi \cdot (h[i-1] - \mu),\, i > 0
\end{align*}
The target function is the log of the joint probability density function (up to a constant)
and we wish to differentiate it with respect to $h_{std}, h, \phi, \sigma, \mu$.
Fig.~\ref{fig:stochastic_volatility} shows the benchmark results.

FastAD outperforms Adept by $2$ times and Stan by $ 2.6$ times for the largest $N$.
The trend seems to stabilize starting from $N = 2^{10}$.
It is interesting to see that FastAD actually performs better than the baseline
for moderate to large $N$ values.
For a complicated model as such, there are many opportunities for FastAD to cache certain
evaluations for constants as mentioned in Section~\ref{ssec:compile-time-opt} 
and~\ref{ssec:normal_log_pdf}.
In particular, the exponential function $ e^h$ reuses its forward-evaluated result,
and many log-pdfs cache the log of its parameters such as 
$\log(\sigma)$ in the Normal log-pdfs 
and $\log(\gamma)$ in the Cauchy log-pdfs 
($\sigma, \gamma$ are the second parameters of their respective distributions, 
which are constant in this model).
Note that this caching is automatically done in FastAD, 
which would be tedious to manually code for the baseline.
Hence, this shows that due to automatic caching, 
FastAD forward and backward-evaluation combined 
can be faster than a manually-written forward evaluation only, 
which puts FastAD at an optimal performance.

\section{Conclusion}

In this paper, we first introduced the reverse-mode automatic differentiation algorithm
to give context and background on how FastAD is implemented.
We then discussed how FastAD uses vectorization to boost the overall performance, 
expression template and lazy allocation strategies to simplify memory management,
and compile-time checks to further reduce run-time.\
To see how FastAD performs in practice, we rigorously benchmarked 
a set of micro and macro-benchmarks
with other popular C++ AD libraries and showed that FastAD consistently achieved
2 to 19 times faster speed than the next two fastest libraries across a wide range of examples.

\section{Further Studies}

FastAD is still quite new and 
does not have full support for all commonly-used functions,
probability density functions,
and matrix decompositions.
While FastAD is currently optimized for CPU performance,
its design can also be extended to support GPU.\@
Having support for both processors will allow FastAD
to be well-suited for extremely large-scale problems as well.
Although computing hessian would be easy to implement in FastAD,
generalizing to higher-order derivatives seems to pose a great challenge,
and this feature could be useful in areas such as physics and optimization problems.
Lastly, one application that could potentially benefit greatly from FastAD 
is probabilistic programming language such as Stan,
which heavily uses automatic differentiation for differentiating scalar functions.

\section{Acknowledgements}

We would like to give special thanks to the members of the Laplace Lab at Columbia University,
especially Bob Carpenter, Ben Bales, Charles Margossian, and Steve Bronder,
as well as Andrea Walther, Eric Phipps, and Brad Bell,
the maintainers of ADOL-C, Sacado, and CppAD, respectively,
for their helpful comments on the first draft and 
for taking the time to optimize the benchmark code for their respective libraries.
We also thank Art Owen and our colleagues Kevin Guo, Dean Deng, and John Cherian
for useful feedback and corrections on the first draft.

\bibliographystyle{splncs04}
\bibliography{references}

\end{document}